\documentclass[trackchanges]{aa}  

\usepackage{graphicx}
\usepackage{txfonts}
\begin{document}

   \title{Deriving the bulk properties of solar wind electrons observed by Solar Orbiter}   

   \subtitle{A preliminary study of electron plasma thermodynamics}

   \author{G. Nicolaou
          \inst{1,2}
          \and
          R. T. Wicks\inst{2,3}
          \and
          C. J. Owen\inst{2}
          \and
          D. O. Kataria\inst{2}
          \and
          A. Chandrasekhar\inst{2}
          \and
          G. R. Lewis\inst{2}
          \and
          D. Verscharen\inst{2,4}
          \and
          V. Fortunato\inst{5}
          \and
          G. Mele\inst{6}
          \and
          R. DeMarco\inst{7}
          \and
          R. Bruno\inst{7}
          }

   \institute{Southwest Research Institute,
              San Antonio, TX  78238, USA\\
              \email{georgios.nicolaou@swri.org}
         \and
             Department of Space and Climate Physics, Mullard Space Science Laboratory, University College London, Dorking, Surrey RH5 6NT, UK
         \and
             Department of Mathematics, Physics and Electrical Engineering, Northumbria University, Newcastle, UK
         \and
             Space Science Center, University of New Hampshire, Durham NH 03824, USA
         \and
             Planetek, Via Massaua, 12, 70132 Bari BA, Italy
         \and
             Leonardo, Taranto, Italy
         \and
             National Institute for Astrophysics, Institute for Space Astrophysics and Planetology, Via del Fosso del Cavaliere 100, I-00133 Roma, Italy
             }

   \date{Received day month year; accepted day month year}

 
  \abstract
   {We demonstrate the calculation of solar wind electron bulk parameters from recent observations by Solar Wind Analyser Electron Analyser System on board Solar Orbiter. We use our methods to derive the electron bulk parameters in a time interval of a few hours. We attempt a preliminary examination of the polytropic behavior of the electrons by analyzing the derived electron density and temperature. Moreover, we discuss the challenges in analyzing the observations due to the spacecraft charging and photo-electron contamination in the energy range < 10 eV.}
   {We derive bulk parameters of thermal solar wind electrons by analyzing Solar Orbiter observations and we investigate if there is any typical polytropic model that applies to the electron density and temperature fluctuations.}
   {We use the appropriate transformations to convert the observations to velocity distribution functions in the instrument frame. We then derive the electron bulk parameters by a) calculating the statistical moments of the constructed velocity distribution functions and b) by fitting the constructed distributions with analytical expressions. We firstly test our methods by applying them to an artificial data-set, which we produce by using the forward modeling technique.}
   {The forward model validates the analysis techniques which we use to derive the electron bulk parameters. The calculation of the statistical moments and the fitting method determines bulk parameters that are identical within uncertainty to the input parameters we use to simulate the plasma electrons in the first place. An application of our analysis technique to the data reveals a nearly isothermal electron "core". The results are affected by the spacecraft potential and the photo-electron contamination, which we need to characterize in detail in future analyses.}
   {}


   \maketitle
%
\section{Introduction}

The ESA Solar Orbiter mission investigates the solar wind dynamics within the inner heliosphere. The scientific payload consists of four in-situ and six remote sensing instruments. 
The Solar Wind Analyser instrument suite \citep[SWA,][]{Owen2020} is one of Solar Orbiter's in-situ instruments and comprises three sensors which share a 
 common data processing unit. SWA measurements resolve the velocity distribution functions (VDFs) of solar wind protons, $\alpha$-particles, electrons and heavier ions in every few seconds. 
 The orbit of the spacecraft and the ability of SWA to measure the VDFs of the solar-wind particles with such a high time resolution, 
 gives us the great opportunity to study physical mechanisms within a broad range of time-scales (even at kinetic scales) in the inner heliosphere \citep[e.g.,][]{Verscharen2019,Zouganelis2020}. 

SWA sensors are designed to operate in several modes. In normal mode operations for example, the instruments resolve the three-dimensional (3D) VDFs of the solar wind particles, while in burst mode the measurements construct 2D pitch angle distributions that capture the shortest time-scale processes. However, the scientific interpretation of the observations, often requires sophisticated analysis tools and methods. Sometimes, a model of the instrument's response (i.e. a forward model) is the main tool to derive the plasma bulk parameters, such as density, bulk flow velocity and temperature \citep[e.g.][]{Nicolaou2014, Nicolaou2015MT, Nicolaou2019, Nicolaou2020a,Wilson2008,Wilson2017,Cara2017,Elliott2016}. For instance, \cite{Nicolaou2014} use a forward model of the Solar Wind Around Pluto instrument \citep[SWAP,][]{McComas2008} on board New Horizons to fit the count distributions observed by the instrument in the deep Jovian magnetosheath. The results quantify the dynamical motions of the Jovian magnetotail and basic thermodynamic properties of the plasma ions. \citet{Elliott2016,Elliott2019} use a forward model of the same instrument and derive the plasma properties in the outer heliosphere. The derived properties reveal how the solar wind evolves as it propagates through the heliosphere. 

The development of such models is not just extremely useful for the analysis of the observations, but also for the quantification of the errors in the derived parameters which define the confidence level of the scientific results \citep{Nicolaou2020c,Criton2020}. The basic principle is to simulate the observations of a specific instrument in given plasma conditions. The modeling requires the knowledge of the instrument's characteristics which are determined from the final hardware design and its detailed calibration. We can then analyze the simulated measurements in a way we plan to analyze the actual observations. Finally, we compare the analysis results with the input plasma parameters we use to simulate the observations in the first place. The comparison quantifies the capabilities of the instrument and the analysis methods in resolving specific features of the plasma VDFs. For example, the study by \cite{Nicolaou2018} demonstrates and quantifies the accuracy of a novel method deriving the kappa indices of proton VDFs constructed from measurements by SWA's Proton Alpha Sensor (SWA-PAS). With the use of a forward model, the authors show that the accuracy depends on the plasma parameters in a complicated way.  

Here, we develop a model of SWA's Electron Analyser System (SWA-EAS) which is designed to resolve the VDFs of solar wind plasma electrons \citep{Owen2020}. Typical solar wind electron VDFs consist of three electron populations: i) the thermal "core", which comprises the majority of the electron density, ii) the nearly isotropic "halo" which extends to higher energies and ii) the field aligned beams, known as the "strahl" electrons \citep[e.g.,][]{Feldman1975,Pilipp1987}. In this paper, we use our model to produce the expected normal mode observations for typical "core" electrons, taking into account the initial calibration of the instrument's energy range and resolution, angular range and resolution, and detection efficiency. In section \ref{instrument_section}, we describe briefly the SWA-EAS instrument and in section \ref{model_section}, we describe how we model the VDFs in each of the two SWA-EAS sensor heads, how we convert the observations onto the instrument frame and analyze them. In section \ref{results_section}, we show modeled observations of typical thermal solar wind electrons, with their velocities characterized by idealized VDFs, which we analyze in order to validate the necessary coordinate transformations that construct the 3D VDFs in the instrument frame. In section \ref{application_section}, we apply our analysis tools to a time interval of SWA-EAS observations and we derive the plasma bulk properties that allow us to study the thermodynamic behavior of the thermal plasma electrons. More specifically, we investigate the relationship between the electron density and temperature in order to derive the polytropic index, a useful parameter for energy transfer studies. In section \ref{discussion_section}, we discuss the results of this work and identify several challenges regarding the spacecraft potential and photo-electron contamination, which we need to overcome in future analyses. Finally, section \ref{conclusions_section} summarizes our conclusions.

\section{SWA-EAS instrument} \label{instrument_section}
\begin{figure}
   \centering
   \includegraphics[angle=0,width=8cm]{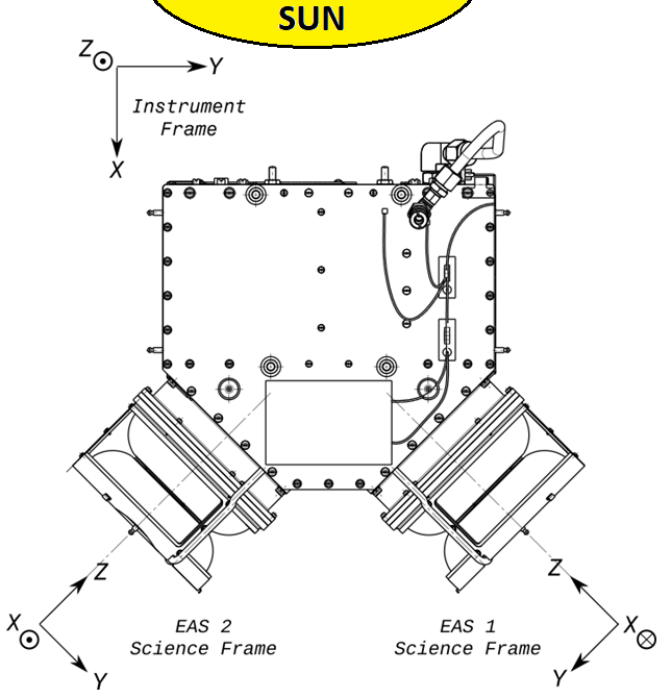}
   \caption{Diagram of the instrument's reference frame and the reference frame of the two SWA-EAS heads. The anti-sunward direction is along the x-axis of the instrument frame. The two SWA-EAS heads are mounted orthogonally on the electronics box which is installed on the spacecraft boom in the shadow of the spacecraft.}
    \label{instrument_schematic}
   \end{figure}

\begin{figure*}
   \centering
   \includegraphics[width=18cm]{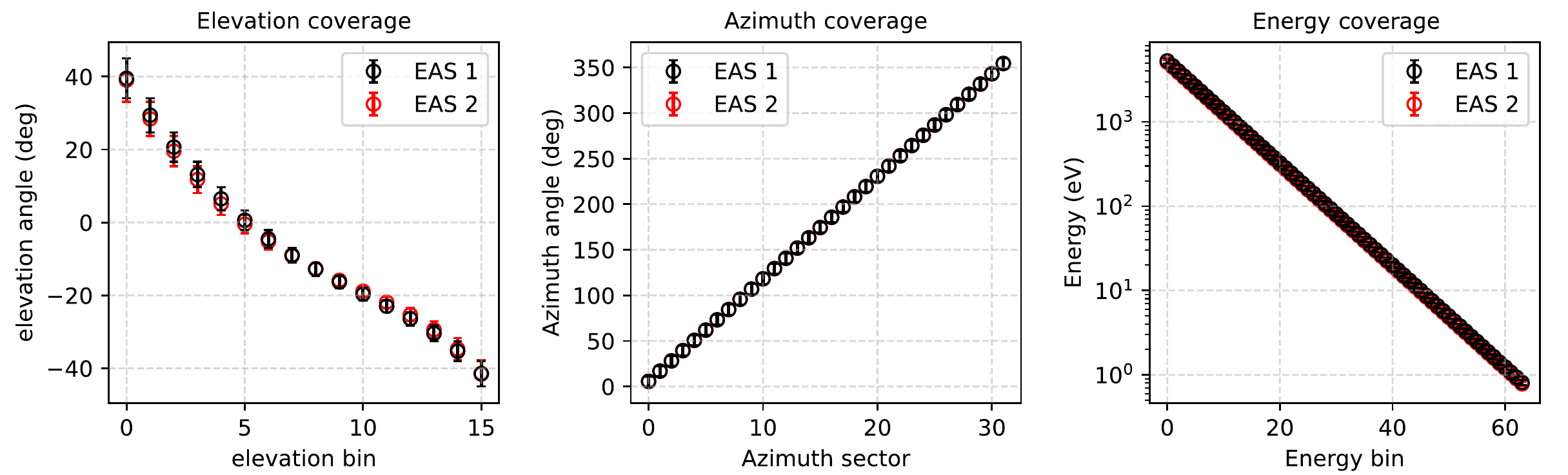}
   \caption{(Left) the elevation angle, (middle) azimuth angle, and (right) energy coverage for (black) SWA-EAS 1 and (red) SWA-EAS 2 heads. In each panel, the circles indicate the center of each bin, while the error bars correspond to their bandwidth.}
    \label{coverage}
   \end{figure*}

SWA-EAS is designed to measure the 3D VDFs of solar wind electrons at heliocentric distances between $\sim$0.3 and $\sim$1 au. In order to achieve full-sky observations, the instrument comprises two top-hat electrostatic analyzer (ESA) heads; SWA-EAS 1 and SWA-EAS 2. The two heads are mounted orthogonally on an electronics box installed on the main spacecraft boom in the shadow of the spacecraft (see Figure \ref{instrument_schematic}). The reference frame of each SWA-EAS head has its z-axis perpendicular to the top-hat plane, pointing towards the box. The x-axis of SWA-EAS 1 points southwards while the x-axis of SWA-EAS 2 points northwards. The y-axis of each SWA-EAS head frame completes a right handed orthogonal frame. The instrument frame has its x-axis along the anti-sunward direction and its z-axis northwards. Each analyzer scans the elevation angle $\Theta$ of the incoming electrons in its own frame, which is defined as the angle between the particle velocity vector and the top-hat plane, increasing towards negative z-axis. The measured elevations and the elevation range are slightly different for each SWA-EAS head, but roughly, each head covers the range between $-45^{\circ}$ and $+45^{\circ}$, in 16 steps of the aperture deflectors. The elevation bandwidth $\Delta\Theta$ varies with the elevation angle and it ranges from $\sim 3^{\circ}$ to $\sim12^{\circ}$ for both sensors (see left panel of Figure \ref{coverage}). The azimuth direction of the incoming particles, for each SWA-EAS head, is defined as the angle between the particle velocity projection on the x-y plane (top-hat plane) and the x-axis in the corresponding SWA-EAS frame. Each SWA-EAS resolves the azimuth direction $\Phi$ in each energy-elevation scan simultaneously, by using 32 azimuth sectors mounted on the position sensitive Micro Channel Plate (MCP) detector which is parallel to the top-hat plane. The azimuth sectors in each SWA-EAS head cover the full 2$\pi$ range. Each azimuth sector has a bandwidth $\Delta\Phi \sim$11.25$^{\circ}$. Finally, each SWA-EAS measures the energy $E$ of the plasma electrons in the range between $\sim$0.7 eV and 5 keV in 64 log-spaced energy steps. The energy bandwidth of each energy step is $\Delta E\,/\,E$ $\sim 12.5\%$. The left panel of Figure \ref{coverage} shows the elevation coverage of the instrument ($\Theta \pm \Delta\Theta/2$), the middle panel shows the azimuth angle coverage ($\Phi \pm \Delta\Phi/2$), while the right panel shows the energy coverage ($E \pm \Delta E/2$).

\section{Model and Methods}\label{model_section}
\subsection{VDF model}
We first model the solar wind electrons with their velocities following the isotropic Maxwellian distribution functions:
   \begin{equation}
   f(\vec{u})=n(\pi v_{\mathrm{th}}^{2})^{-\frac{3}{2}}\mathrm{exp}\left[ -\frac{(\vec{u}-\vec{v})^2}{v_{\mathrm{th}}^2} \right],
   \label{Maxwellian}
   \end{equation}
\noindent
where \vec{u} is the particle velocity vector, $n$ is the electron density, $v_{\mathrm{th}}$ is the thermal speed, and $\vec{v}$ is the bulk flow velocity vector of the plasma electrons. We define our input distribution function, including the bulk velocity vector in the instrument reference frame.

 \subsection{VDF in SWA-EAS heads} \label{rotation_matrices}
In order to model the distributions constructed from the observations of each EAS head, we calculate the velocity vector in each EAS head frame by applying rotation matrices. The velocity vector components in the SWA-EAS 1 frame are

\begin{equation}
   \begin{bmatrix} 
    v_{\mathrm{x,EAS1}} \\ 
    v_{\mathrm{y,EAS1}}  \\
    v_{\mathrm{z,EAS1}}
    \end{bmatrix}=
    \begin{bmatrix}
    0 & 0 & -1 \\ 
    \mathrm{sin}(45^{\circ}) & -\mathrm{cos}(45^{\circ}) & 0  \\
    -\mathrm{cos}(45^{\circ}) & -\mathrm{sin}(45^{\circ}) & 0
    \end{bmatrix}
    \begin{bmatrix}
    v_{\mathrm{x}} \\ 
    v_{\mathrm{y}}  \\
    v_{\mathrm{z}}
    \end{bmatrix}
\label{EAS1_conversion}
 \end{equation}

\noindent
Similarly, the velocity components in the SWA-EAS 2 are

\begin{equation} 
   \begin{bmatrix} 
   v_{\mathrm{x,EAS2}} \\ 
   v_{\mathrm{y,EAS2}}  \\
   v_{\mathrm{z,EAS2}}
   \end{bmatrix}
   = 
   \begin{bmatrix} 
   0 & 0 & 1 \\ 
   \mathrm{sin}(45^{\circ}) & \mathrm{cos}(45^{\circ}) & 0  \\
   -\mathrm{cos}(45^{\circ}) & \mathrm{sin}(45^{\circ}) & 0
   \end{bmatrix}
   \begin{bmatrix} 
   v_{\mathrm{x}} \\ 
   v_{\mathrm{y}}  \\
   v_{\mathrm{z}}
   \end{bmatrix}
\label{EAS2_conversion}
\end{equation}

\begin{figure*}
   \centering
   \includegraphics[width=18cm]{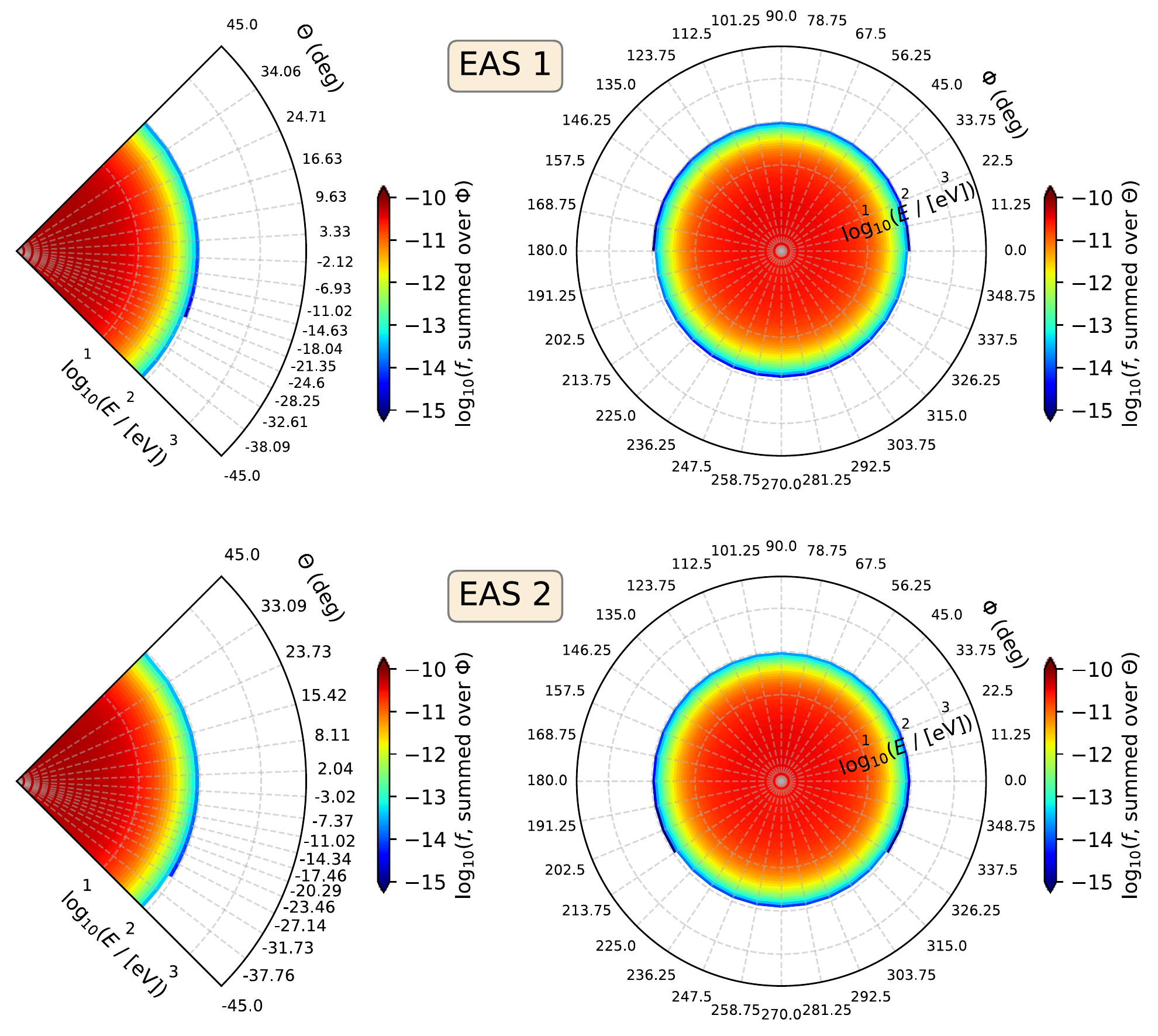}
   \caption{Model distribution function constructed in the (top) SWA-EAS 1 and (bottom) SWA-EAS 2 head. For this model we use input density $n$ = 100 cm$^{-3}$, bulk velocity $\vec{v}$ = 440 km$\,\mathrm{s}^{-1}$ and direction along the x-axis of the instrument frame, and $v_{\mathrm{th}}$ = 2000 km$\,\mathrm{s}^{-1}$. In the left panel of each row, we show the model distribution function as a function of energy and elevation angle, summed over azimuth. The right panels show the model distribution as a function of energy and azimuth, summed over elevation.}
    \label{EAS1_EAS2_distribution}
   \end{figure*}

We model the response of each SWA-EAS head, considering their energy and field of view range and resolution. We take into account that each SWA-EAS head resolves the distribution in an $E$, $\Theta$, $\Phi$ grid. We further assume that the constructed VDF in each $E$, $\Theta$, $\Phi$ pixel has the value corresponding to the exact center of the pixel. This simplification ignores the shape of the VDF and a possibly non-uniform response of the instrument within the bandwidth of each pixel. In Figure \ref{EAS1_EAS2_distribution}, we show an example of a distribution modeled on (top panel) SWA-EAS 1 and (bottom panel) SWA-EAS 2. For this example we use the input parameters $n$ = 100 cm$^{-3}$, $\vec{v}$= 440$\hat{x}$ km$\,\mathrm{s}^{-1}$, and $v_{\mathrm{th}}$ = 2000 km$\,\mathrm{s}^{-1}$.

  \subsection{Conversion to instrument frame} 
\begin{figure*}
   \centering
   \includegraphics[angle=0,width=18cm]{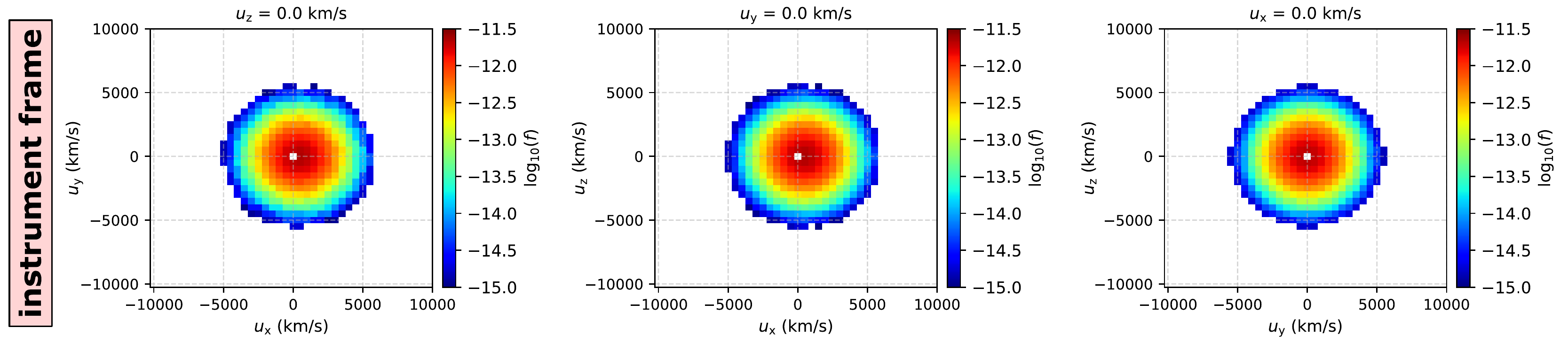}
   \caption{2D cuts of the model distribution function shown in Figure \ref{EAS1_EAS2_distribution}, constructed in the instrument frame. (Left) the distribution function f at $u_{\mathrm{z}}$ = 0 km$\,\mathrm{s}^{-1}$ as a function of $u_{\mathrm{x}}$ and $u_{\mathrm{y}}$, (middle) $f$ at $u_{\mathrm{y}}$ = 0 km$\,\mathrm{s}^{-1}$ as a function of $u_{\mathrm{x}}$ and $u_{\mathrm{z}}$, and (right) $f$ at $u_{\mathrm{x}}$ = 0 km$\,\mathrm{s}^{-1}$ and as a function of $u_{\mathrm{y}}$ and $u_{\mathrm{z}}$.}
    \label{Instrument_frame_cuts}
   \end{figure*}

  Each SWA-EAS head constructs part of the electron VDF from the observations. The first step of the analysis combines the observations by both SWA-EAS heads and determines the full 3D VDF in the instrument frame. We do that by defining a grid on the instrument frame and applying the rotation matrices presented in Equations \ref{EAS1_conversion} and \ref{EAS2_conversion} in order to project each pixel of the instrument frame to the SWA-EAS 1 and SWA-EAS 2 frames respectively. Then, each projected pixel assumes the VDF value constructed for the specific $E$, $\Theta$, $\Phi$ pixel of an individual head that includes the projected point within the bandwidth $\Delta E$, $\Delta \Theta$, $\Delta \Phi$. For pixels that are covered by both heads, we apply the average of the two VDF values corresponding to each head's value in that bin. In Figure \ref{Instrument_frame_cuts}, we show 2D cuts of the model distribution function shown in Figure \ref{EAS1_EAS2_distribution}, as constructed in the instrument frame with resolution $\Delta u_{\mathrm{x}} \times \Delta u_{\mathrm{y}} \times \Delta u_{\mathrm{z}}$ = $500\,\times\,500\,\times\,500\,\mathrm{km^{3}\,s^{-3}}$. In principle, we can define the grid of the instrument frame as we wish in order to either reduce the computation steps of our analysis or to improve the accuracy of our analysis. The accuracy however, is bounded by the instrument resolution and cannot increase indefinitely. Ideally, the grid definition should be optimized for specific applications, depending on the shape of the constructed VDFs.  

After the construction of the 3D VDFs in the instrument frame, we calculate the bulk parameters of the electrons, using two typical analysis methods. We firstly calculate the plasma bulk parameters by numerically calculating the statistical moments of the constructed velocity distribution functions. The second method fits the constructed distribution functions with analytical expressions by optimizing the electron bulk parameters. 

\subsection{Velocity Moments} \label{moments_section}
 We firstly calculate the velocity moments of the electrons by numerically integrating (summing discrete pixels) the constructed velocity distribution functions. For the constructed $f(u_{\mathrm{x}},u_{\mathrm{y}},u_{\mathrm{z}})$ determined in the instrument frame, the plasma density is given by

   \begin{equation}
   n_{\mathrm{mom}}=\sum_{u_\mathrm{x}}\sum_{u_\mathrm{y}}\sum_{u_\mathrm{z}} f(u_{\mathrm{x}},u_{\mathrm{y}},u_{\mathrm{z}}) \mathrm{\Delta}u_{\mathrm{x}}\,\mathrm{\Delta}u_{\mathrm{y}}\,\mathrm{\Delta}u_{\mathrm{z}}.
   \end{equation}

\noindent
The bulk velocity components are

   \begin{equation}
   v_{\mathrm{x,mom}}=\frac{\sum_{u_\mathrm{x}}\sum_{u_\mathrm{y}}\sum_{u_\mathrm{z}} u_{\mathrm{x}} f(u_{\mathrm{x}},u_{\mathrm{y}},u_{\mathrm{z}}) \mathrm{\Delta}u_{\mathrm{x}}\,\mathrm{\Delta}u_{\mathrm{y}}\,\mathrm{\Delta}u_{\mathrm{z}}}{n_{\mathrm{mom}}},
   \end{equation}
   
      \begin{equation}
   v_{\mathrm{y,mom}}=\frac{\sum_{u_\mathrm{x}}\sum_{u_\mathrm{y}}\sum_{u_\mathrm{z}} u_{\mathrm{y}} f(u_{\mathrm{x}},u_{\mathrm{y}},u_{\mathrm{z}}) \mathrm{\Delta}u_{\mathrm{x}}\,\mathrm{\Delta}u_{\mathrm{y}}\,\mathrm{\Delta}u_{\mathrm{z}}}{n_{\mathrm{mom}}},
   \end{equation}
 \noindent and
      \begin{equation}
   v_{\mathrm{z,mom}}=\frac{\sum_{u_\mathrm{x}}\sum_{u_\mathrm{y}}\sum_{u_\mathrm{z}} u_{\mathrm{z}} f(u_{\mathrm{x}},u_{\mathrm{y}},u_{\mathrm{z}}) \mathrm{\Delta}u_{\mathrm{x}}\,\mathrm{\Delta}u_{\mathrm{y}}\,\mathrm{\Delta}u_{\mathrm{z}}}{n_{\mathrm{mom}}}.
   \end{equation}

\noindent
The thermal speed is calculated as
    \begin{equation}\label{eq_Vth}
    v_{\mathrm{th,mom}}=\sqrt{ \frac{\frac{2}{3}\sum_{u_\mathrm{x}}\sum_{u_\mathrm{y}}\sum_{u_\mathrm{z}} [(\vec{u}-\vec{v_{\mathrm{mom}}})^2] f(u_{\mathrm{x}},u_{\mathrm{y}},u_{\mathrm{z}}) \mathrm{\Delta}u_{\mathrm{x}}\,\mathrm{\Delta}u_{\mathrm{y}}\,\mathrm{\Delta}u_{\mathrm{z}}}{n_{\mathrm{mom}}}} .
   \end{equation}

\subsection{Fitting} \label{fitting_section}
The fitting analysis is widely used to determine the analytical form of observed distribution functions. The fitting method optimizes the parameters of an analytical expression for $f(u_{\mathrm{x}},u_{\mathrm{y}},u_{\mathrm{z}})$, which we define as $f_{\mathrm{fit}}(u_{\mathrm{x}},u_{\mathrm{y}},u_{\mathrm{z}})$. Typically, a fitting routine minimizes the chi-square parameter $\chi^{2}$, which in its simplest form is

    \begin{equation}\label{fit_equation}
    \chi^2 = \sum_{u_{\mathrm{x}}}\sum_{u_{\mathrm{y}}}\sum_{u_{\mathrm{z}}}[f(u_{\mathrm{x}},u_{\mathrm{y}},u_{\mathrm{z}})-f_{\mathrm{fit}}(u_{\mathrm{x}},u_{\mathrm{y}},u_{\mathrm{z}})]^2,
   \end{equation}
\noindent
where we use Equation \ref{Maxwellian} as $f_{\mathrm{fit}}$ \footnote{Occasionally, it is preferable to fit the logarithm of the constructed VDFs, as this approach captures better the tails of the distribution functions \citep[e.g.,][]{Nicolaou2018,Nicolaou2020a}. However, through this paper we use Equation \ref{fit_equation} as we focus on the analysis of the "core" electrons.}, parameterized by the bulk parameters we want to estimate:

   \begin{equation}
   f_{\mathrm{fit}}(\vec{u})=n_{\mathrm{fit}}(\pi v_{\mathrm{th,fit}}^{2})^{-\frac{3}{2}}\mathrm{exp}\left[ -\frac{(\vec{u}-\vec{v_{\mathrm{fit}}})^2}{v_{\mathrm{th,fit}}^2} \right].
   \label{Maxwellian_fit}
   \end{equation}

\section{Model Results}\label{results_section}
\begin{figure}
   \centering
   \includegraphics[angle=0,width=9cm]{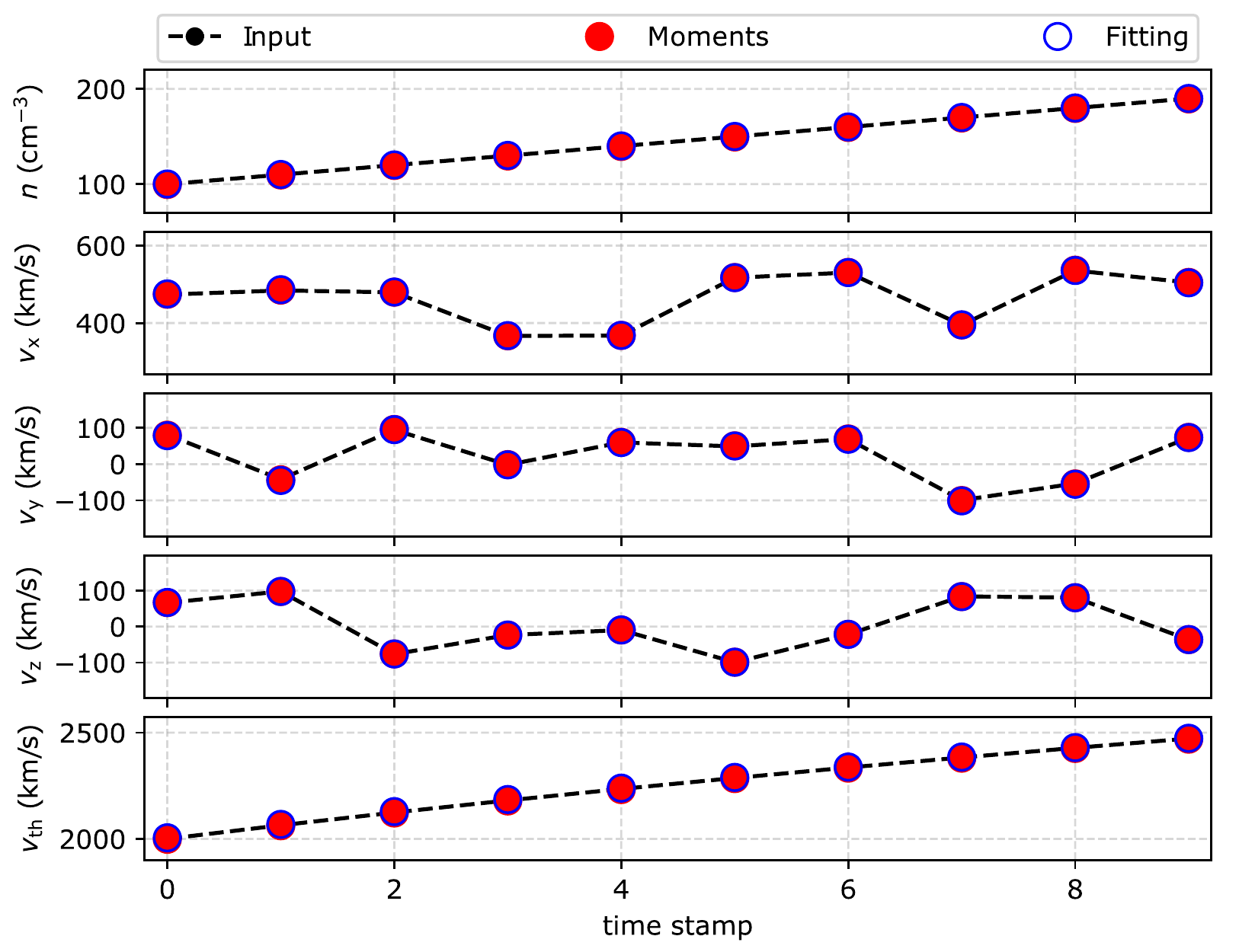}
   \caption{Time series of (black) input, (red) derived by moments and (blue) derived by fitting plasma bulk parameters of "core" electrons with their velocities following isotropic Maxwell distribution functions. There is not an apparent difference between the derived and input parameters and thus, our analysis is appropriate for this plasma parameter range.}
    \label{model_time_series}
   \end{figure}

We first test our analysis tools by using the forward model to simulate ten VDFs of solar wind electrons with different bulk parameters. We consider plasma electrons with their density increasing from 100 to 190 cm$^{-3}$ (see Figure \ref{model_time_series}). We set random fluctuations of the velocity vector components with an amplitude $\delta v_{\mathrm{x}}$ = $\delta v_{\mathrm{y}}$ = $\delta v_{\mathrm{z}}$ = 100 km$\,\mathrm{s}^{-1}$ and average values $v_{\mathrm{x}}$ = 440 km$\, \mathrm{s}^{-1}$, $v_{\mathrm{y}}$ = $v_{\mathrm{z}}$= 0 km$\,\mathrm{s}^{-1}$. Finally, the thermal speed increases with the plasma density according to the adiabatic relationship (i.e. $v_{\mathrm{th}}\propto n^{1/3}$), which is widely used to describe space plasmas within the heliosphere \citep[e.g.,][]{Parker1961,Zhang2016}. We then construct the VDFs in the Cartesian instrument frame with resolution $\Delta u_{\mathrm{x}} \times \Delta u_{\mathrm{y}} \times \Delta u_{\mathrm{z}}$ = $500\,\times\,500\,\times\,500\,\mathrm{km^{3}\,s^{-3}}$ and analyze them as explained in Sections \ref{moments_section} and \ref{fitting_section} to derive the electron bulk parameters. 

In Figure \ref{model_time_series}, we show the time series of the input and the derived parameters. The derived parameters are almost identical to the corresponding input parameters, indicating that the VDF transformation to the instrument frame and the analysis tools we are using, are appropriate.    

\begin{figure*}
   \centering
   \includegraphics[angle=0,width=18cm]{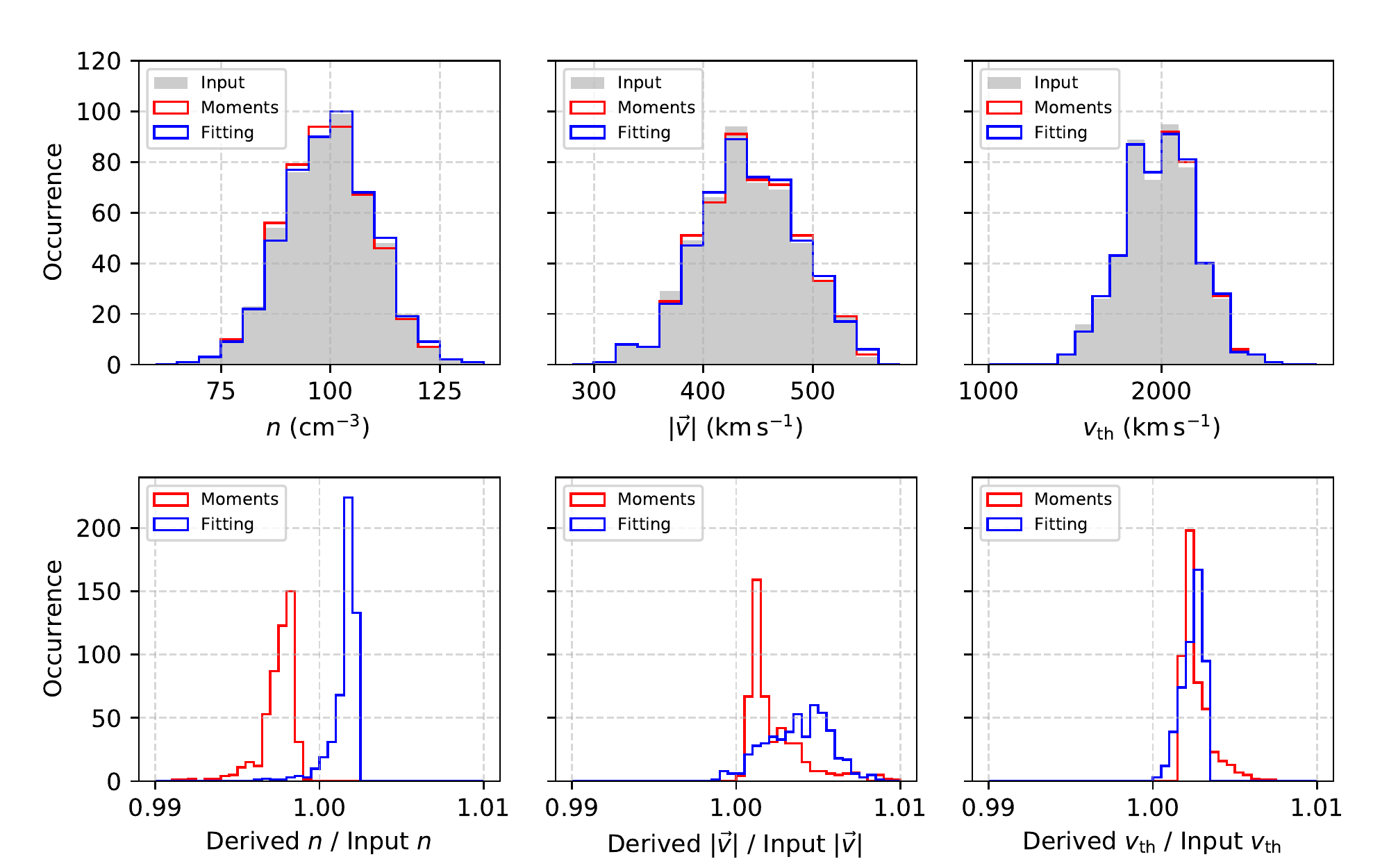}
   \caption{(Top left) Histogram of (gray) input, (red) derived by moments, and (blue) derived by fit plasma density. (Top middle) Histogram of the input and derived plasma speed and (top right) the input and derived thermal speed in the same format. (Bottom left) Histogram of (red) derived by moments and (blue) derived by fitting density divided by the input density. (Bottom middle) Histogram of the derived over input speed and (bottom right) the derived over input thermal speed in the same format.}
    \label{model_histograms}
   \end{figure*}

In order to examine the accuracy of our analysis tools for a wider range of plasma parameters, we perform an additional test. We use the forward model to simulate the observations of 500 Maxwellian VDFs, with their bulk parameters randomly selected from normal distributions. Specifically, we simulate 500 observations with an average plasma density $\bar{n}$ = 100 cm$^{-3}$ and standard deviation $\sigma{\bar{n}}$ = 10 cm$^{-3}$, average velocity components $\bar{v}_{\mathrm{x}}$ = 440 km$\,\mathrm{s}^{-1}$,  $\bar{v}_{\mathrm{y}}$ = $\bar{v}_{\mathrm{z}}$ = 0 km$\,\mathrm{s}^{-1}$ with standard deviations $\sigma{v_{\mathrm{x}}}$ = $\sigma{v_{\mathrm{y}}}$ = $\sigma{v_{\mathrm{z}}}$ = 50 km$\,\mathrm{s}^{-1}$ respectively, and an average thermal speed $\bar{v}_{\mathrm{th}}$ = 2000 km$\,\mathrm{s}^{-1}$ with standard deviation $\sigma{v_{\mathrm{th}}}$ = 200 km$\,\mathrm{s}^{-1}$. We then analyze the 500 modeled VDFs and compare the results with the input plasma parameters. The top panels of Figure \ref{model_histograms} show histograms of the 500 input parameters $n$, |$\vec{v}$|, and $v_{\mathrm{th}}$ and the corresponding parameters derived by the moments and fitting analysis. The input bulk parameters and the corresponding parameters derived by both analysis methods have similar histograms. We quantify the agreement between the analysis results and the actual parameters by studying histograms of the ratios of the derived parameters over the corresponding input parameters (see bottom panels of Figure \ref{model_histograms}). The moments analysis of our idealized input distributions derives more accurately the parameters than the fitting analysis for the examined range of input parameters. Nevertheless, the average values of the ratios in both analysis methods are remarkably close to unity (within 0.4$\%$) and the standard deviations are less than 0.2$\%$. Our results show that we can have high confidence in our methods we use to re-sample and analyze EAS measurements.

\section{Application to flight data} \label{application_section}
\subsection{Time series of bulk parameters}
We apply our analysis methods to a set of SWA-EAS measurements obtained on 15-06-2020 from 14:50:19UT to 17:35:19UT, when the spacecraft was at $\sim$0.52 au from the Sun. We first convert the electron counts $C(E,\Theta,\Phi)$ to electron VDFs in each SWA-EAS head. For the conversion we use the simplified formula \citep[e.g.,][]{Franz2006,Lavraud2016}

   \begin{equation}
   f(E,\Theta,\Phi)=\frac{m_{\mathrm{e}}^2C(E,\Theta,\Phi)}{2E^2G_{\mathrm{f}}(\Theta,\Phi)Q_{\mathrm{e}}(E)\Delta\tau},
   \label{Maxwellian_fit}
   \end{equation}
\noindent
where $m_{\mathrm{e}}$ is the electron mass, $G_{\mathrm{f}}(\Theta,\Phi)$ is the geometric factor of the electrostatic analyzer head as determined in ground calibration, $Q_{\mathrm{e}}(E)$ is the quantum efficiency of the detector, also as determined in ground calibration, and $\Delta\tau$ is the acquisition time for each $E,\,\Theta$ scan. The $G_{\mathrm{f}}$ is independent of energy, while the quantum efficiency depends only on the energy of the detected particles. For nominal mode operations, $\Delta\tau\sim$ 0.85 milliseconds.  

\begin{figure}
   \centering
   \includegraphics[angle=0,width=9cm]{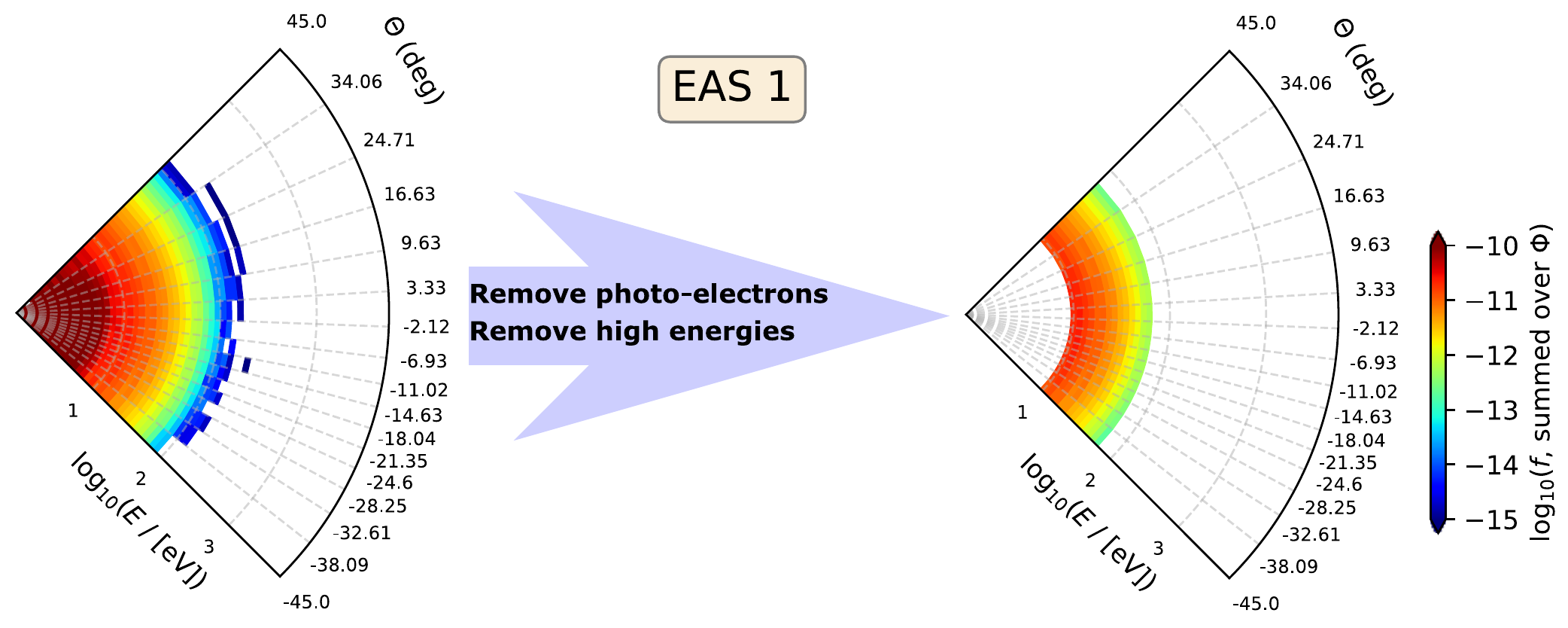}
   \caption{(Left) The electron VDF as a function of log$_{10}(E)$ and $\Theta$, summed over $\Phi$,  constructed on the SWA-EAS 1 head from observations obtained at 14:50:19UT on 15-06-2020. (Right) the same distribution with all electrons with energies below $\sim$10 eV and above $\sim$68 eV removed.}
    \label{cleaning}
   \end{figure}

We want to characterize the part of the VDF that contains the thermal "core" electrons. Therefore, we exclude data-points in energy channels $E \, > \, $ 68 eV, which is about the typical energy range of supra-thermal electrons and electron beams \citep[e.g.,][]{Feldman1975,Maksimovic2005}. Moreover, the lower energy range of the constructed distribution is contaminated by photo-electrons which are produced on the spacecraft body, mainly by UV radiation. These photo-electrons are accelerated by the spacecraft potential. Our preliminary investigation of the data-set, shows that the photo-electron count distributions have their peaks below 10 eV (see Appendix \ref{appendix1}). Therefore, in this first attempt to analyze the data, we exclude the measurements in energy channels $E \, < \, $ 10 eV. In the left panel of Figure \ref{cleaning}, we show an example of a measured electron VDF as a function of log$_{10}(E)$ and $\Theta$, summed over $\Phi$,  constructed in the EAS 1 head. In the right panel of Figure \ref{cleaning}, we show the VDF part with energies between $\sim$ 10 and $\sim$ 68 eV which is the VDF part we analyze.

We also find a systematic difference between the VDFs constructed by the two EAS heads. We apply a preliminary correction in order to eliminate the differences in the measurements by EAS 1 and EAS 2 while observing the same VDF. The details of the sensitivity cross-calibration are given in Appendix \ref{appendix2}. We note however, that the two sensors will be cross calibrated in the near future. The detailed in-flight calibration will be based on the analysis of the entire data-set obtained up to this time and will provide detailed efficiency corrections.

\begin{figure*}
   \centering
   \includegraphics[angle=0,width=18.5cm]{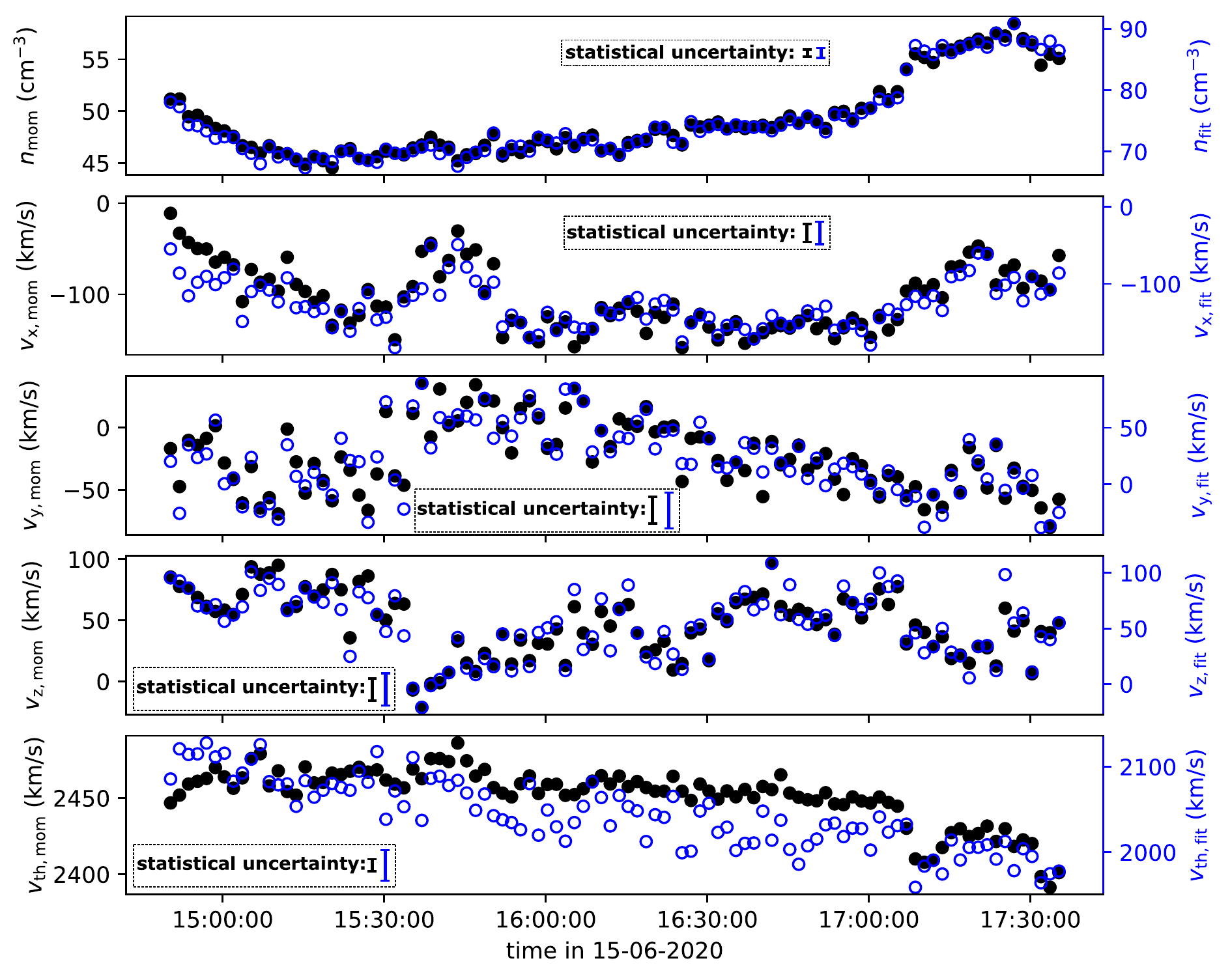}
   \caption{Bulk parameters of solar wind electrons derived by (black) moments analysis and (blue) fitting analysis of the velocity distribution functions constructed from observations obtained between 14:50:19UT and 17:35:19UT on 15-06-2020. From top to bottom, we show the electron density, velocity along x-direction, velocity along y-direction, velocity along z-direction, and thermal speed. In each panel, we show the typical statistical uncertainty for the derived parameters.}
    \label{Fit_Moments_time_series}
   \end{figure*}

After the energy range selection and the scaling of $f_{\mathrm{EAS2}}$, we construct the VDF in the instrument frame and apply the moments and the fitting analysis to derive the plasma electron bulk parameters. Figure \ref{Fit_Moments_time_series} shows the results of this analysis. We also calculate and show the typical statistical uncertainty of the derived parameters. Our calculations assume that the observed counts follow the Poisson distribution (see Appendix \ref{appendix3}). The electron density derived by moments $n_{\mathrm{mom}}$ is significantly smaller than the density derived by fitting $n_{\mathrm{fit}}$ within the entire interval. This is an expected result since we exclude a significant portion (i.e., the part outside the energy range between 10 and 68 eV) of the analyzed VDF and the derived $n_{\mathrm{mom}}$ is actually a partial density. The electron velocity components derived by both methods follow a similar pattern. However, the absolute values are different, especially for the $v_{\mathrm{y}}$ component. Finally, the thermal speeds derived by both methods follow a very similar pattern, with $v_{\mathrm{th,mom}}$ being $\sim 20\%$ larger than $v_{\mathrm{th,fit}}$. This difference is also expected considering that the formula deriving $v_{\mathrm{th,mom}}$ has the partial (underestimated) density in the denominator (Equation \ref{eq_Vth}). In Section \ref{discussion_section}, we discuss further the expected differences between the derived parameters.  

We quantify the linear relationships and the linear correlation coefficients between the parameters derived by moments and the corresponding parameters derived by fitting. In Table 1, we show the results of this analysis. The two densities $n_{\mathrm{mom}}$ and $n_{\mathrm{fit}}$ are almost perfectly correlated (Pearson coefficient $\sim$ 0.99). All plasma velocity components have a strong correlation (Pearson coefficient > 0.9). The correlation between the derived thermal speeds is smaller, but still significant ($\sim$0.78).

%
\begin{table*} \label{Cor_table}
\caption{Linear relationships and correlation coefficients between the bulk parameters derived by moments and derived by fitting analysis. }             
\label{table}      
\centering                          
\begin{tabular}{ l l l}        
\hline\hline                 
Parameter & Linear Relationship & Correlation \\    
\hline                        
   $n$ & $n_{\mathrm{fit}}=1.72n_{\mathrm{mom}}-9.73$ (cm$^{-3}$)& 0.99 \\      
   $v_{\mathrm{x}}$ & $v_{\mathrm{x,fit}}=0.80v_{\mathrm{x,mom}}-42.17$ (km$\,\mathrm{s^{-1}}$)& 0.93   \\
   $v_{\mathrm{y}}$ & $v_{\mathrm{y,fit}}=1.02v_{\mathrm{y,mom}}+46.52$ (km$\,\mathrm{s^{-1}}$)& 0.93      \\
   $v_{\mathrm{z}}$ & $v_{\mathrm{z,fit}}=1.06v_{\mathrm{z,mom}}+3.40$ (km$\,\mathrm{s^{-1}}$)& 0.94    \\
   $v_{\mathrm{th}}$ & $v_{\mathrm{th,fit}}=1.70v_{\mathrm{th,mom}}-2135.85$ (km$\,\mathrm{s^{-1}}$)& 0.78    \\ 
\hline                                   
\end{tabular}
\end{table*}
%

\subsection{Comparison with RPW densities}
\begin{figure}
   \centering
   \includegraphics[angle=0,width=9cm]{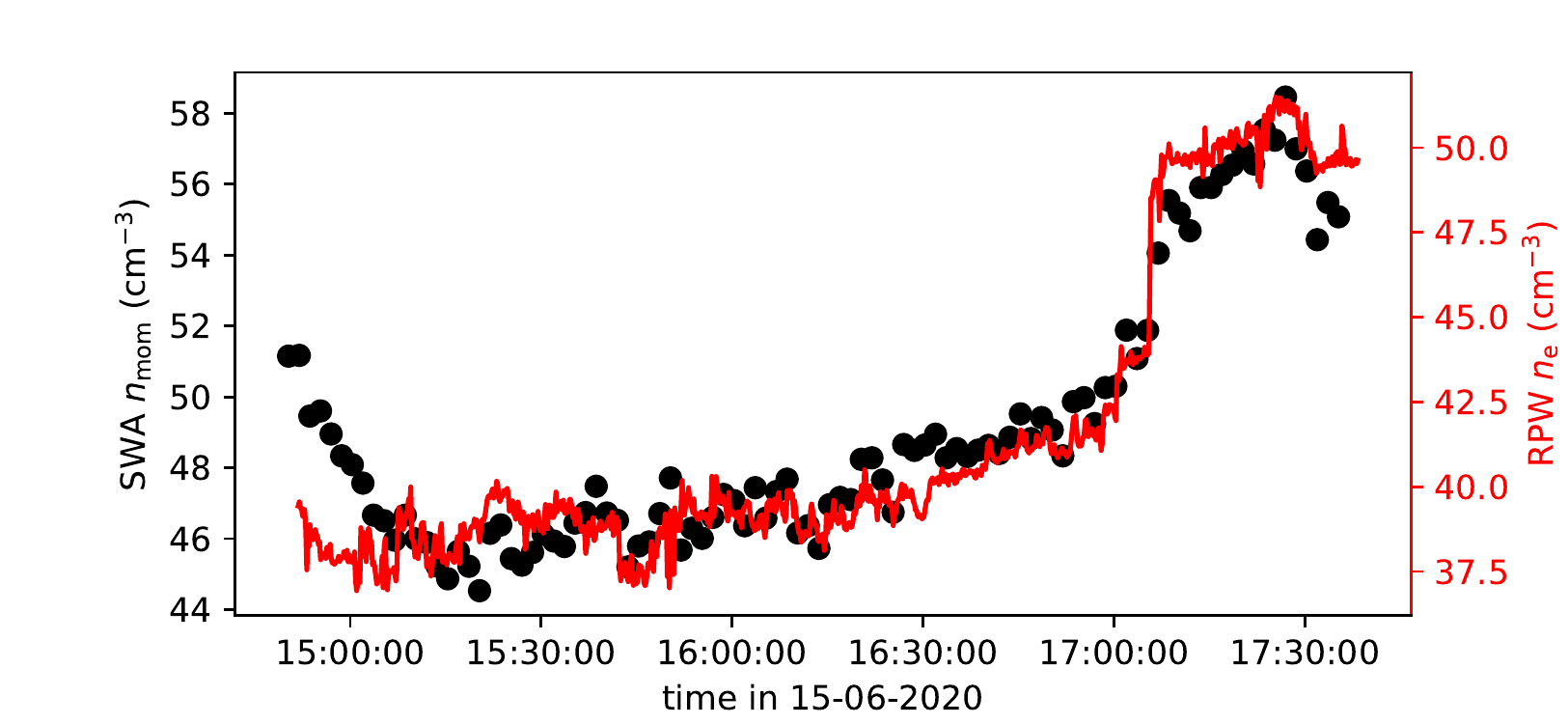}
   \caption{Electron plasma density (black) as derived here from the SWA-EAS observations analysis and (red) as observed by RPW instrument on-board.}
    \label{RPW_comparison}
   \end{figure}

We compare the electron density values we derive from EAS-SWA observations with the electron density values derived by the Solar Orbiter Radio and Plasma Waves (RPW) instrument \citep{Maksimovic2020} on board Solar Orbiter, for the same time interval (Figure \ref{RPW_comparison}). Although there is a systematical offset between the derived values, the two instruments observe a similar pattern of the density variations for the most of the time interval we examine. This is of course an encouraging result, showing the analysis of data from two different instruments with a completely different measurement principle, determine similar dynamic variations of the electron density.

\subsection{Electron thermodynamics}
One of Solar Orbiter's goals is the understanding of physical mechanisms which involve energy transfer between the plasma species and electromagnetic fields \citep{Zouganelis2020}. Such processes may occur on several time-scales giving characteristic signatures in the VDFs of some or all the plasma species. The investigation of plasma processes involving energy transfer is sometimes achieved through the determination of the polytropic behavior of the plasma. The polytropic process is the transition of a plasma from one thermodynamic state to another following a specific relationship between the density $n$ and temperature $T$ \citep[e.g.,][]{Parker1961}:
   \begin{equation}
   T \propto  n^{\gamma-1},
   \label{polytropic_eq}
   \end{equation}
\noindent
where $\gamma$ is the polytropic index which governs the process. For the special case with $\gamma$ = 1, the plasma has a constant temperature and the process is called isothermal. An isobaric process (constant pressure $P\propto nT$) has $\gamma$ =0, while an isochoric process (constant $n$) has $\gamma$ = $\infty$. Finally, in an adiabatic process, there is no energy exchange and the polytropic index is equal to the ratio of the specific heats $\gamma = c_{\mathrm{p}}/c_{\mathrm{v}}$. The polytropic equation achieves a closure to the VDF moments hierarchy through the relationship between higher order moments (e.g., temperature, pressure) and the zeroth order moment; the plasma density. Polytropic closures are basic ingredients for fluid descriptions of plasmas. However, recent studies show that the fluid description also applies to short-scale fluctuations in the solar wind \citep[e.g.,][]{Wu2019,Verscharen2019}. Typical analyses determine $\gamma$ from the linear relationship between the logarithms of $n$ and $T$. From Equation \ref{polytropic_eq} we get
   \begin{equation}
   \mathrm{log_{10}}(T)=(\gamma-1)\mathrm{log_{10}}(n)+\mathrm{const}.
   \label{polytropic_eq_log}
   \end{equation}
 
For example, several studies fit Equation \ref{polytropic_eq_log} to observations in order to derive the polytropic behavior of solar wind and its structures in the inner heliosphere \citep[e.g.,][]{Totten1995,Bavassano1996,Newbury1997,Nicolaou2020apj}, at $\sim$ 1 au \citep[e.g.,][]{Osherovich1993,Kartalev2006,Nicolaou2014Sol,Livadiotis_Desai2016,Livadiotis2018Entr, Livadiotis2018JGR} and in the outer heliosphere \citep[e.g.,][]{Livadiotis2013,Elliott2019}. Some other studies use the same method to derive the polytropic behavior of the plasma within Earth’s magnetosphere \citep[e.g.,][]{Pang2016,Park2019}, Jovian magnetosheath and boundary layer \citep[e.g.,][]{Nicolaou2014,Nicolaou2015BL}, Saturnian magnetosphere \citep[e.g.,][]{Arridge2009,Dialynas2018}, and more.

\begin{figure}
   \centering
   \includegraphics[angle=0,width=9cm]{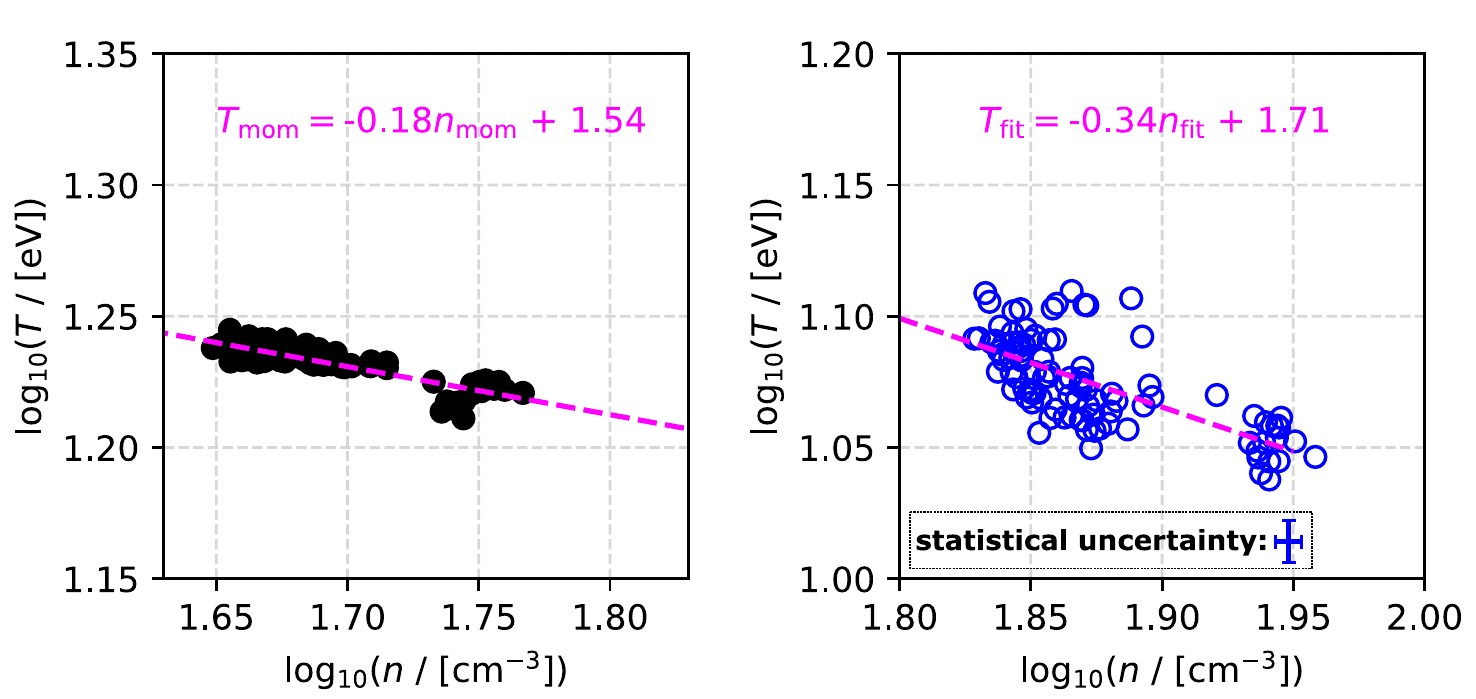}
   \caption{log$_{10}T$ vs. $log_{10}n$ of solar wind electrons, as derived by (left) statistical moments analysis and (right) fitting analysis of the first 100 3D-VDFs obtained in 15-06-2020. The magenta dashed lines in each plot are isobaric lines. There is a clear anti-correlation between log$_{10}(T)$ and log$_{10}(n)$. However, the small value of the slope describing the relationship between the observables, indicates a nearly isothermal electron plasma. The statistical error bar in the left panel is not shown as it is comparable to the data symbol size.}
    \label{polytropic}
   \end{figure}

We examine if the polytropic model applies to the electron bulk parameters we derive here from SWA-EAS observations. In Figure \ref{polytropic}, we show the log$_{10}$($T$) vs. log$_{10}$($n$) for the time interval shown in Figure \ref{Fit_Moments_time_series}. The left panel shows the parameters derived by the moments calculation, and the right panel the parameters derived by fitting. The density and the temperature are anti-correlated. We calculate a Pearson correlation coefficient $\sim$ -0.87 for the moment parameters and $\sim$ -0.69 for the fitting parameters. In both panels, the magenta line is the linear model fitted to the data-set. The calculated slopes indicate $\gamma \sim$ 0.82 when the moments are used and $\gamma \sim$ 0.66 when the fitting parameters are used. These values show a polytropic behavior within the isothermal ($\gamma$ = 1) and isobaric ($\gamma$ = 0) range. According to this result, if the electrons have three effective kinetic degrees of freedom, then there is an energy transfer associated with their bulk parameter fluctuations. It is important to note that proper polytropic behavior investigations should be carefully applied to identified streamlines \citep[e.g.,][]{Kartalev2006,Nicolaou2014Sol,Pang2016,Livadiotis2018Entr,Livadiotis2018JGR,Nicolaou2019Pol,Livadiotis2021,Nicolaou2021}. The identification of streamlines demands analysis of plasma and magnetic field data. We plan such a detailed analysis in the near future, when the cross calibrated data from SWA-EAS are available. In the next section we discuss further our results.  

\section{Discussion}\label{discussion_section}

Our SWA-EAS response model validates the VDF transformation to the instrument frame and confirms the accurate derivation of the electron plasma bulk parameters comparing both a moments calculation and fitting analysis. The application of our methods to flight data shows that the statistical velocity moments and the Maxwellian fits to the analyzed VDFs derive parameters with similar patterns, but different absolute values. 

Firstly, the zeroth-order velocity moments derive smaller densities than the densities derived by the Maxwellian fits to the same VDFs. This is an expected result, considering that we exclude a big portion of the VDFs from our analysis. By excluding measurements in energies < 10 eV which are contaminated by photo-electrons, we also exclude the actual thermal solar-wind electrons at these energies. Therefore, the derived moments are actually partial moments. On the other hand, the Maxwellian fitting interpolates the missing VDFs in the low and high energy range, accounting for the missing parts assuming a Maxwellian VDF. Importantly, besides a systematical offset, the densities we derive compare fairly well with the densities derived by the RPW instrument.

The bulk velocity components derived by moments are strongly correlated with the corresponding velocity components derived by fitting (Pearson coefficient >0.9). However, occasionally, the absolute values exhibit large differences (> 50 km$\,\mathrm{s}^{-1}$). The $v_{\mathrm{x}}$ velocity component is almost always negative within the analyzed data-set, reaching a minimum value -150 km$\,\mathrm{s}^{-1}$. Both the $v_{\mathrm{y}}$ and $v_{\mathrm{z}}$ components fluctuate roughly between -100 and +100 km$\,\mathrm{s}^{-1}$. A negative $v_{\mathrm{x}}$ component indicates a bulk flow towards the Sun, which is the opposite direction of the expected solar wind bulk flow. However, we acknowledge that our instrument does not resolve energies below $\sim$ 0.7 eV, which correspond to electron speeds < 500 km$\,\mathrm{s}^{-1}$. Additionally, here we analyze only the measurements obtained at energies > 10 eV, in order to exclude photo electrons. The energy resolution of the instrument $\Delta E\,/\,E \sim$ 12.5 $\%$, corresponds to a speed resolution $\Delta u\,/\,u=\Delta E\,/\,2E\, \sim$ 6.25 $\%$. This means that even in the lowest energy bin we analyze here (10eV), we measure electrons with $u$ = $\sqrt{2E/m}$ $\sim$1800 km$\,\mathrm{s}^{-1}$ and the speed bandwidth is as large as $\Delta u\,/\,u$ = 118 km$\,\mathrm{s}^{-1}$, which is a significant fraction of the expected solar wind speed. In fact, the peaks of the core electron VDFs which describe best the electron bulk speed are well beyond the analyzed energy range. In order to get an estimation of how the erroneous flow direction affects the derivation of the other bulk parameters, we simulate a VDF with bulk parameters that are typical within the time interval we examine; $n$ = 75 cm$^{-3}$, a pure sunward flow $V_{\mathrm{x}}$ = -150 km$\,\mathrm{s}^{-1}$ and $V_{\mathrm{th}}$ = 2050 km$\,\mathrm{s}^{-1}$. We fit the simulated VDF with a drifting Maxwellian function with a pure antisunward flow $V_{\mathrm{x}}$ = 440 km$\,\mathrm{s}^-1$, which is the expected solar wind flow. The fit derives the plasma density and temperature within 5$\%$ and 10$\%$ of their values, respectively.

\begin{figure}
   \centering
   \includegraphics[angle=0,width=9cm]{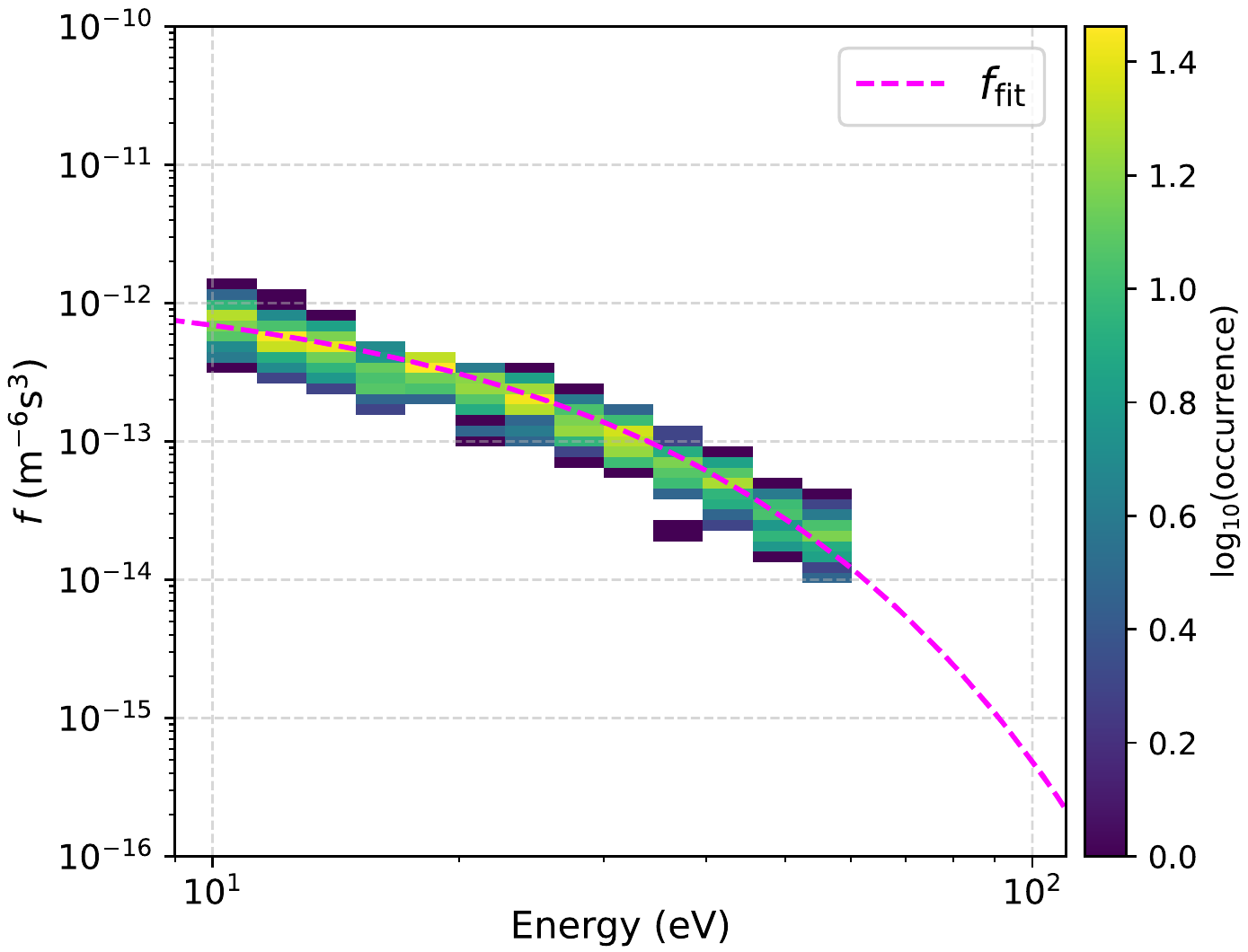}
   \caption{The VDF of plasma electrons as a function of energy, observed by SWA-EAS at 14:50:19UT on 15-06-2020, averaged over elevation and azimuth directions. The dashed-magenta shows the corresponding Maxwellian fit to the observations.}
    \label{fitting_1D}
   \end{figure}

The thermal speed derived by the moments analysis is linearly correlated but systematically larger than the thermal speed derived by the fitting analysis. The difference between the two is of the order of 400 km$\,\mathrm{s}^{-1}$. The main reason for this difference is again the fact that the derived moments are derived from the analysis of partial VDFs (not the entire energy range). According to Equation \ref{eq_Vth}, the derived thermal speed is proportional to $n_{\mathrm{mom}}^{-1/2}$. Therefore, the underestimation of $n_{\mathrm{mom}}$ due to the excluded lower energy range, results in an overestimation of $v_{\mathrm{th,mom}}$. Other differences between $v_{\mathrm{th,mom}}$ and $v_{\mathrm{th,fit}}$ are possibly due to deviations of the constructed VDFs from the Maxwell distribution function. For instance, if the observed VDFs exhibit non-Maxwellian high energy tails, the fitting analysis underestimates the derived temperature \citep[e.g.,][]{Nicolaou2016,Livadiotis2018JGR}. In general, the data we analyze here do not exhibit large deviations from the Maxwellian model, as we show in the typical 1D-fitting example in Figure \ref{fitting_1D}.   

By using either the derived moments or the fitting parameters, we find a similar polytropic model for the plasma electrons (Figure \ref{polytropic}). This result verifies that the moment and the fitting parameters capture fluctuations in $n$ and $T$ in a comparable way, besides the significant differences of their absolute-value offsets for the reasons discussed earlier in this section. According to our analysis, the polytropic model that describes our results has $\gamma \sim$ 0.82 when we use the derived moments or $\gamma \sim$ 0.66 when we use the fitting parameters. These values are close to the isothermal model which is consistent with energy transfer during the bulk fluctuations of plasma particles with three effective kinetic degrees of freedom. A large heat flux in the electrons may lead to the quasi-isothermal behavior of thermal solar wind electrons \citep[e.g.,][]{Hollweg1976}.

We remind the reader that this study characterizes only the electrons in the core of the VDF without applying sophisticated corrections for the photo-electron contamination and the spacecraft potential which are expected to affect the determination of the VDFs \citep[e.g.,][]{Song1997,Salem2001}. Although, we exclude low energy electrons (<10 eV) from our analysis, we expect that a few volts of spacecraft potential can affect the accuracy of our derivations. The spacecraft potential modifies the energies and the directions of the particles \citep[e.g.,][]{Lewis2008,Lewis2010,Lavraud2016,Voshchepynets2018,Bergman2020,Bergman2021}. Detailed studies of the spacecraft potential and its geometry should be prioritized in the future to improve the accuracy of the electron bulk properties.

\section{Summary and Conclusions}\label{conclusions_section}
We analyze the first measurements by the Solar Wind Analyser's Electron Analyser System (SWA-EAS) on board Solar Orbiter. For the evaluation of the analysis method we developed a forward model of the instrument's response. We derive the electron plasma bulk parameters by using two classic analysis methods; calculation of the statistical velocity moments of the constructed velocity distribution functions and fitting the velocity distribution functions with analytical expressions. We finally investigate if there is a general polytropic behavior supported by the analyzed data-set. In summary, we conclude that
   \begin{enumerate}
      \item Our method successfully converts the observations by both SWA-EAS heads to velocity distribution functions in the instrument frame; 
      \item Our analysis derives accurate electron plasma bulk parameters for a typical solar wind plasma; 
      \item Low energy photo-electrons contaminate the observations within the low energy range  <10 eV. By excluding the contaminated parts of the analyzed distributions we underestimated the density moment and overestimate the thermal speed moment;
      \item The spacecraft potential is comparable to the energy of the electrons in the "core" of the velocity distribution function. Therefore the spacecraft potential is expected to manipulate the energies and the directions of the solar wind electrons just before they get detected by SWA-EAS. A thorough study of the spacecraft potential is required in order to correct the energies and the directions of the observed electrons.
     \item The fitting process described here is able to improve some of the calculated moment inaccuracies, i.e., the fitted density is almost two times higher because the fit can compensate for the lost counts in the low energy range and so calculates the temperature more accurately also. There is a remaining issue, likely caused by the spacecraft photo-electrons accelerated by the spacecraft potential, that is making  the velocity calculated from both moments and fits to be significantly modified from the expected solar wind, to the extend that the radial velocity is pointing in the wrong direction.
     \item This first analysis of the derived bulk parameters, suggest that the observed fluctuations of the thermal electrons have a large heat flux, leading to a quasi-isothermal variability.
   \end{enumerate}
Our future plan is to fully characterize the resolved 3D VDFs from the fully calibrated data and derive their bulk properties. We will then investigate the polytropic behavior of the entire solar wind electron population (including supra-thermal electrons and electron beams if possible) and compare with the polytropic behavior of solar wind protons and heavier ions observed by Solar Wind Analyser's Proton Alpha Sensor (SWA-PAS) and Heavy Ion Sensor (SWA-HIS) respectively \citep{Owen2020}. 

\begin{acknowledgements}
      G.N., R.T.W., C.J.O. and D.V. are supported by the STFC Consolidated Grant to UCL/MSSL, ST/S000240/1. D.V. is supported by STFC Ernest Rutherford Fellowship ST/P003826/1.
\end{acknowledgements}

%
%

%
\bibliographystyle{aa}
\bibliography{references}

\begin{appendix}
\section{Photo-electron Contamination}\label{appendix1} 
Photo-electrons are produced on the spacecraft body and are accelerated by the spacecraft potential. If the paths of the accelerated photo-electrons are within the range of the sampled directions, we expect to detect them at energies that are comparable with the spacecraft potential. The distributions of the observed photo-electrons overlap with the VDFs of solar wind electrons, making it very difficult to resolve the actual solar wind VDFs within the lower energy range. In this first analysis attempt, we detect the energy range of the majority of the photo-electrons and we exclude all the measurements obtained within this energy range.

\begin{figure}
   \centering
   \includegraphics[angle=0,width=9cm]{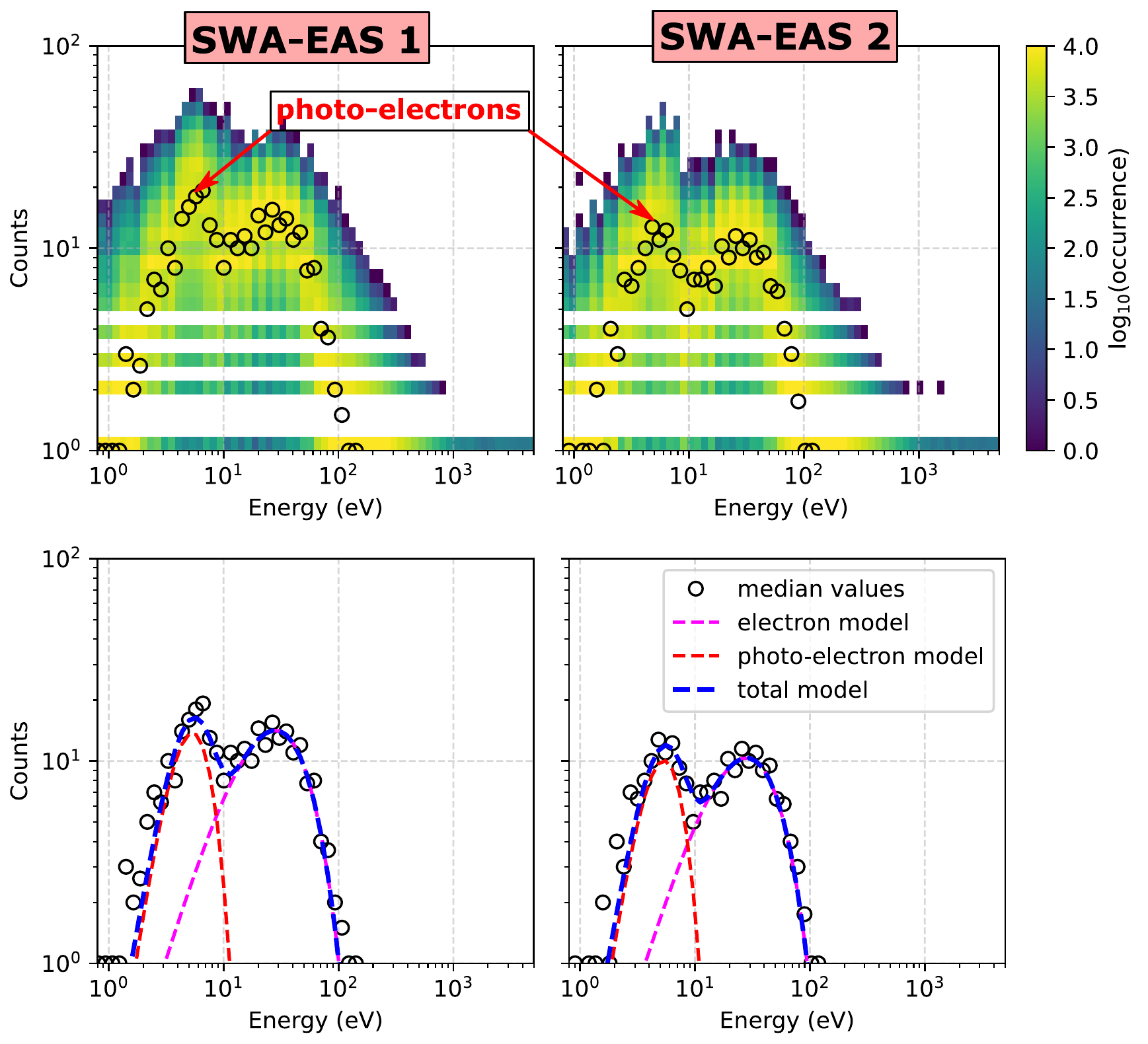}
   \caption{(Top left) Distribution of counts as a function of energy for the observations obtained from 14:50:19UT to 17:35:19UT on 15-06-2020, by EAS 1 head and (top right) by EAS 2 head. Each bin of the histogram shows the sum of the counts within the time interval, elevation bins and azimuth sectors. The black data-points show the median value of counts in each energy bin. (Bottom left) The median value of counts in each energy bin as measured by EAS 1, and (bottom right) by EAS2. The red and magenta dashed lines show models of the photo-electrons and thermal electrons respectively. Photo-electrons and thermal electrons are well separated 10 eV.}
   \label{photo-electrons}
   \end{figure}

In order to investigate the energy range of the photo-electrons, we study the count distribution as a function of energy within the analyzed interval. The top left panel of Figure \ref{photo-electrons} shows this distribution for the EAS 1 head, while the top right panel shows the same distribution for the EAS 2. The black data-points in both panels show the median values of counts as a function of energy. The bottom panels show and the median value of the counts measured in each energy bin, along with a photo-electron model and a thermal electron model. Both models are based on Maxwellian VDFs. We clearly see two peaks in the count distributions. The lower energy peak corresponds to the photo-electrons, while the second peak corresponds to the thermal electrons. We set 10 eV (red dashed line in Figure \ref{photo-electrons}) as the energy level that separates the photo-electron and thermal electron distributions.

\section{Sensitivity cross-calibration}\label{appendix2}
\begin{figure}
   \centering
   \includegraphics[angle=0,width=9cm]{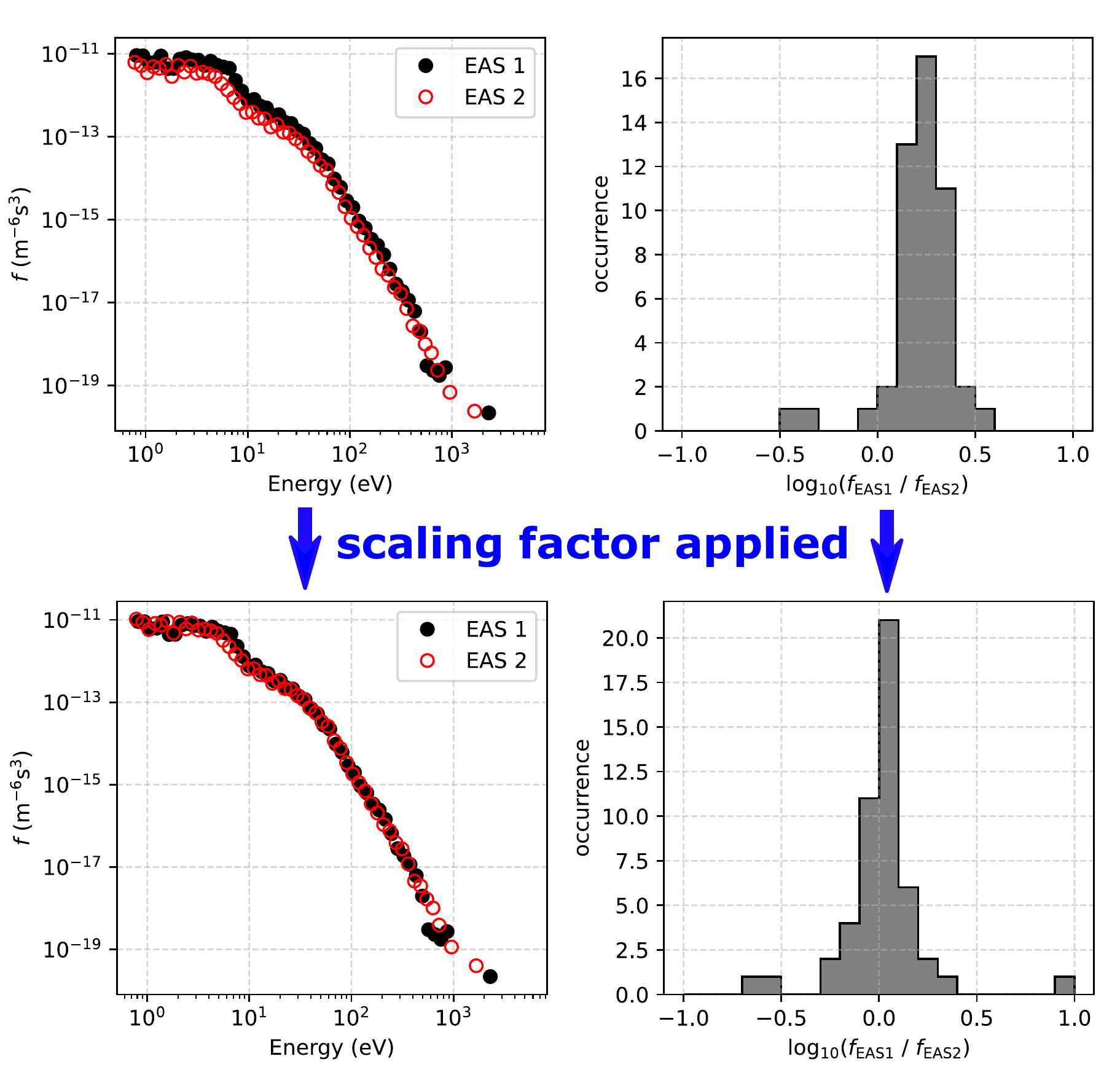}
   \caption{(Top left) The VDF of solar wind electrons as a function of $E$, constructed from the (black) SWA-EAS 1 and (red) SWA-EAS 2 head, by including only the elevation and azimuth bins closest to the +z direction in the instrument frame. The purple shadowed region marks the energy range we analyze in this study. (Top right) Histogram of the ratio values of the VDF constructed from SWA-EAS 1 measurements, divided by the corresponding VDF constructed by SWA-EAS 2 measurements. (Bottom left) The same as in the top left panel, but with SWA-EAS 2 measurements multiplied by 1.65, and (bottom right) shows the corresponding ratio after the multiplication.}
    \label{f1_f2_scaling}
   \end{figure}

 By using the most recent laboratory calibration factors, our analysis reveals systematically larger $f(E,\Theta,\Phi)$ values constructed by EAS 1 than the corresponding values constructed by EAS 2 for the same distribution. Figure \ref{f1_f2_scaling} shows this systematic difference. In  the top panel, the black data points show the VDF as a function of energy, constructed by EAS 1 ($f_{\mathrm{EAS1}}$) in a single elevation bin and azimuth sector, which sample the +z direction in the reference frame of the instrument. The red data-points in the same panel show the corresponding values of the VDF constructed by EAS 2 ($f_{\mathrm{EAS 2}}$) for the same direction. The shadowed region corresponds to the energy range which we analyze in this study. The top-right panel shows a histogram of log$_{10}$($f_{\mathrm{EAS1}}/f_{\mathrm{EAS2}}$). The peak of the distribution is clearly $>$ 0. The bottom panels of Figure \ref{f1_f2_scaling} show the same plots for $f_{\mathrm{EAS 2}}$ multiplied by a factor of 1.65. The scaling factor recovers the differences between the sensitivity of the two SWA-EAS heads during the analyzed time interval. Therefore, we use this scaling factor for our preliminary study. However, we plan a detailed cross calibration for the entire data-set in the near future.

\section{Typical statistical uncertainties}\label{appendix3}
\begin{figure*}
   \centering
   \includegraphics[angle=0,width=19cm]{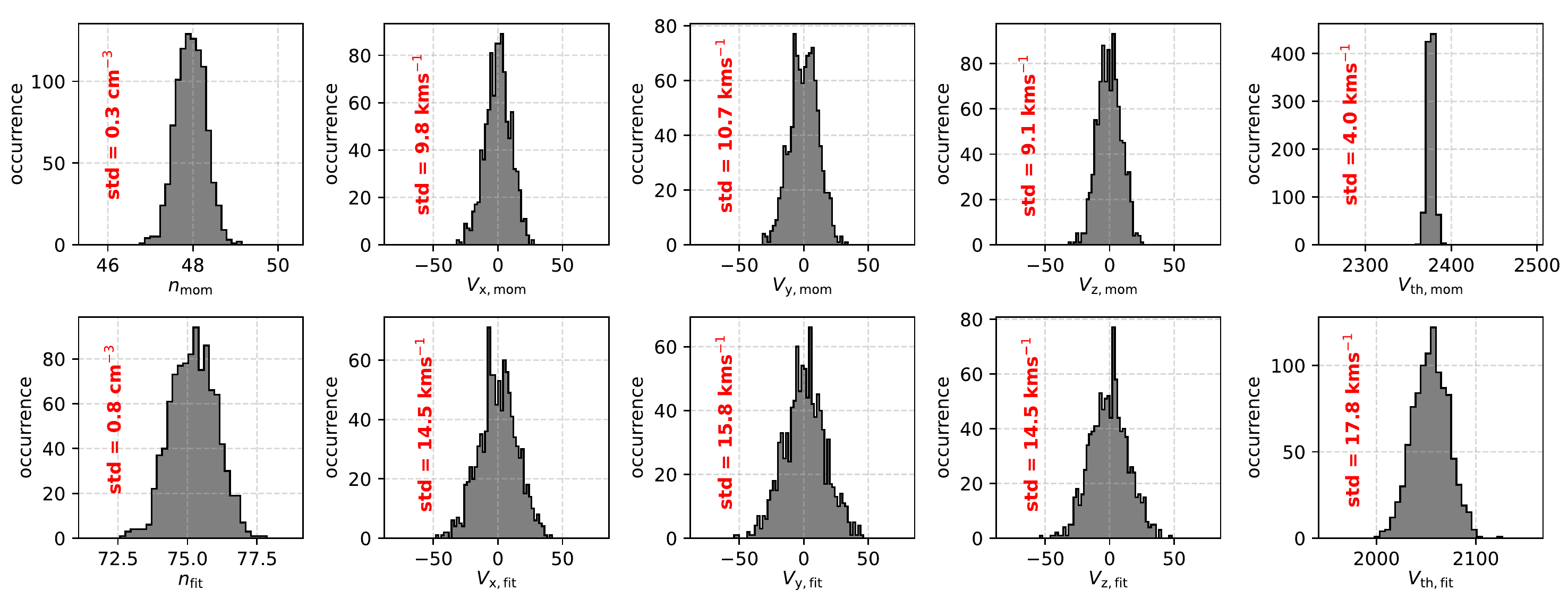}
   \caption{Histograms of the electrons plasma parameters derived by (top) statistical moments and by (bottom) fitting analysis of 1000 simulated plasma samples of plasma with $n$ = 75 cm$^{-3}$ and $V_{\mathrm{th}}$ = 2050 km$\,\mathrm{s}^{-1}$. In our simulations we assume that the number of counts follows the Poisson distribution. In each panel we show the standard deviation of the histogram, which we consider as the typical statistical error for the corresponding parameter. Those are used as error bars in Figures \ref{Fit_Moments_time_series} and \ref{polytropic}.}
    \label{statistical_err_hists}
   \end{figure*}

We estimate the typical statistical uncertainty (measurements error) of the derived plasma parameters, assuming that the number of counts recorded by the instrument follows the Poisson distribution. For our estimations, we simulate 1000 measurement samples of a non-drifting plasma with $n$ = 75 cm$^{-3}$ and $V_{\mathrm{th}}$ = 2050 km$\,\mathrm{s}^{-1}$. This is a typical set of electron plasma parameters within the range of the time series in Figure \ref{Fit_Moments_time_series}. We then analyze the simulated samples the same way we analyze the observations. We show the histograms of the derived parameters in Figure \ref{statistical_err_hists}. The standard deviation of each histogram is then used as the characteristic error bar of the corresponding parameter in Figures \ref{Fit_Moments_time_series} and \ref{polytropic}. Note however that our preliminary analysis uses the initial calibration factors as determined from laboratory testing instrument response models. These initial calibration factors do not account for spacecraft charging effects, or the systematic offset from the response of other instrument's on-board. These corrections are planned for the near future and may reveal additional systematical errors in the derived parameters.

\end{appendix}

\end{document}